\documentclass[11pt]{article}
\usepackage{amssymb}
\usepackage{amsmath}
\usepackage{amsthm}
\usepackage[dvips]{graphics,color}
\usepackage{epsfig}
\usepackage{natbib}

\usepackage{a4wide}

\def\E{\mathbb{E}}
\def\var{\text{Var}}

\def\U{\mathcal{U}}
\newtheorem{thm}{Theorem}

\begin{document}

\title{Fast estimation of multivariate stochastic volatility}

\author{Kostas Triantafyllopoulos\footnote{Department of
Probability and Statistics, Hicks Building, University of Sheffield,
Sheffield S3 7RH, UK, email: {\tt
k.triantafyllopoulos@sheffield.ac.uk}} \and Giovanni
Montana\footnote{Department of Mathematics, Statistics Section,
Imperial College London, London SW7 2AZ, UK, email: {\tt
g.montana@imperial.ac.uk}}}

\maketitle

\begin{abstract}

In this paper we develop a Bayesian procedure for estimating
multivariate stochastic volatility (MSV) using state space models. A
multiplicative model based on inverted Wishart and multivariate
singular beta distributions is proposed for the evolution of the
volatility, and a flexible sequential volatility updating is
employed. Being computationally fast, the resulting estimation
procedure is particularly suitable for on-line forecasting. Three
performance measures are discussed in the context of model
selection: the log-likelihood criterion, the mean of standardized
one-step forecast errors, and sequential Bayes factors. Finally, the
proposed methods are applied to a data set comprising eight exchange
rates \emph{vis-$\grave{a}$-vis} the US dollar.

\textit{Some key words:} multivariate time series, stochastic
volatility, GARCH, state space models, Bayesian forecasting,
Kalman filter, Wishart distribution.

\end{abstract}

\section{Introduction}\label{s1}

Over the last two decades, considerable effort has been devoted to
the development of time-varying volatility models and related
computational algorithms. It is widely recognized that volatility
modeling has important implications for the analysis of returns on
stocks and exchange rates. More recently, attention has moved to
examining the implications of volatility for other financial
applications such as derivatives pricing, optimal portfolio
selection, and risk management (for instance, to enable efficient
forecasting of Value-at-Risk). Although several univariate
volatility models have been developed and are routinely used, the
time-changing feature of the volatility is better described by
multivariate models that explicitly account for cross-correlations
among asset returns. A multivariate framework is desirable because
assets can be formally linked together and can be influenced by
common unobserved factors; as a consequence of this, we often
observe related movements between markets, or sectors, or exchange
rates.

The many efforts to model multivariate volatility fall into two main classes of models: multivariate generalized
auto-regressive heteroscedastic (M-GARCH) models and multivariate stochastic volatility (MSV) models. The review paper
by \cite{Bauwens06} well describes the capabilities and limitations of M-GARCH models. In brief, the large number of
parameters, which are typically specified by maximum likelihood estimation, and the fact that the unobserved
volatility is not modelled as a stochastic process, somehow limit the applicability of these models. On the other
hand, MSV models are more flexible, because the volatility is assumed to change stochastically according to a latent
process. However, most stochastic volatility models, as reviewed for instance in \cite{Yu}, \cite{Liensfield},
\cite{Asai06}, and \cite{Maasoumi06}, need essentially to resort to stochastic simulation schemes such as Markov chain
Monte Carlo methods (MCMC), which may be heavily computationally intensive. Although much progress has been made on
the front of simulation-based procedures, and more efficient algorithms are now available, the iterative nature of
such procedures hampers the applicability of multivariate stochastic volatility estimation in real-time applications
where, for instance, prompt user interventions may be required \citep{Salvador1}. For such reasons, it would be
desirable to rely on analytic solutions that translate into fast and flexible algorithms, while still enjoying some of
the advantages offered by MSV models.

Computational solutions that trade off the complexity of the model for speed are valuable, and have been explored in
the literature. A simplification that facilitates the development of inferential procedures is to assume that the
volatility follows a random walk (RW) evolution. This assumption has been often adopted in the relevant literature,
for instance in the works of \cite{Quintana1,Putnam,Quintana2,West01,Uhlig97,Liu,Soyer06,Carvalho1}, and references
therein. For instance, \cite{Harvey94} suggest an approximate inferential method for a MSV model based on the extended
Kalman filter using crude mean and variance approximations; although the evolution of the volatility matrix is defined
as an autoregressive (AR) process, the authors suggest that a RW evolution works equally well.


In this work we elaborate on some of the results that have already been proposed in the literature mentioned above.
Using the convolution of the Wishart and singular multivariate beta distributions, which was first proved in
\cite{Uhlig94}, we construct a RW model for the evolution of the volatility. In the works of \cite{Aguilar00},
\cite{Liu}, \cite{Soyer06}, and \cite{Carvalho1}, all adopting the RW assumption, the multivariate volatility
estimators resemble their counterpart univariate estimators based on gamma and beta distributions \citep{West01,
Triantafyllopoulos07}. However, we have noticed that these estimators are incorrectly derived, in that they give rise
to a shrinkage volatility evolution, which is not a realistic choice. In particular, we demonstrate how the
multivariate beta density has been overlooked in the above references to the point that the updating equation for the
degrees of freedom has been wrongly computed. The resulting volatility estimator proposed in this paper is a weighted
average of the square logarithmic returns. Thus, with proper choice of the weights, the modeller obtains volatility
estimators that guarantee mean reversion over time and are appropriate to analyze volatility.

This paper is organized as follows. Section \ref{model:model}
defines the model and the Bayesian estimation procedure is given in
Section \ref{model:estimation}. Section \ref{model:performance} is
concerned with model assessment and selection, and three performance
measures are derived, namely the log-likelihood criterion, the mean
of the standardized one-step forecast errors, and sequential Bayes
factors. Section \ref{example:fx} applies our methods to a data set
comprising eight foreign exchange rates \emph{vis-$\grave{a}$-vis}
the US dollar. A proof of Section \ref{model:performance} can be
found in the appendix.

\section{Stochastic volatility}\label{model}

\subsection{The model}\label{model:model}

Consider a $p$-variate vector of log-returns
$\{y_t\}_{t=1,\ldots,N}$, where $t$ is the time index, for some
positive integer $N$. The zero-drift conditional volatility model
assumes
\begin{equation}\label{model1}
y_t=\Sigma_t^{1/2} \epsilon_t, \quad \epsilon_t \sim N_p(0,I_p),
\quad t=1,\ldots,N,
\end{equation}
where $\Sigma_t$ is the conditional volatility matrix of $y_t$,
$\epsilon_t$ is $p$-variate innovation vector following a
$p$-variate Gaussian distribution with zero mean vector and identity
covariance matrix; finally, $\Sigma_t^{1/2}$ denotes the square root
of $\Sigma_t$, using the Choleski decomposition or the spectral
decomposition \citep{Gupta}.

At time $t$, let $y^t=\{y_1,\ldots,y_t\}$ denote the information
set, comprising data up to time $t=1,\ldots,N$. In order to estimate
$\Sigma_t$, we need to define an evolution law for $\Sigma_t$. A
sensible law postulates that
\begin{equation}\label{eq:mean1}
\E(\Sigma_{t+1}^{-1}|y^t)=\E(\Sigma_t^{-1}|y^t),
\end{equation}
namely the expectation from time $t$ to $t+1$ remains unchanged, and
$$\var(\textrm{vecp}(\Sigma_{t+1}^{-1})|y^t) \geq
\var(\textrm{vecp}(\Sigma_t^{-1})|y^t),$$ where
$\textrm{vecp}(\Sigma_t^{-1})$ denotes the column stacking operator
of the covariance matrix $\Sigma_t^{-1}$. These assumptions define a
random-walk type evolution law for $\Sigma_t^{-1}$, i.e.
$\Sigma_{t+1}^{-1}=\Sigma_t^{-1}+\Gamma_t$, where $\Gamma_t$ has
zero mean. Such an evolution is possible under the multiplicative
law of covariance matrices of \cite{Uhlig94}, that is
\begin{equation}\label{model2}
\Sigma_{t+1}^{-1} = k \U (\Sigma_t^{-1})' B_{t+1} \U
(\Sigma_t^{-1}), \quad t=0,1,\ldots,N-1,
\end{equation}
where $\U(\Sigma_t^{-1})$ denotes the upper triangular matrix of the
Choleski decomposition of $\Sigma_t^{-1}$, so that
$\Sigma_t^{-1}=\U(\Sigma_t^{-1})'\U(\Sigma_t^{-1})$. Here $B_{t+1}$
follows, independently of $\Sigma_t^{-1}$, the singular multivariate
beta distribution (whose density is given in equation \eqref{Bt:pdf}
of the appendix). Initially, we assume the inverted Wishart prior
\begin{equation}\label{prior}
\Sigma_0\sim IW_p(n+2p,S_0), \quad n=\frac{1}{1-\delta},
\end{equation}
with density function
$$
p(\Sigma_0)=\frac{ |S_0|^{(n+p-1)/2} \textrm{etr}(-S_0\Sigma_0^{-1})
} { 2^{p(n+p-1)/2} \Gamma_p((n+p-1)/2) |\Sigma_0|^{(n+2p)/2}},
$$
where $0<\delta<1$ is a discount factor, $|S_0|$ is the
determinant of $S_0$, $\textrm{etr}(.)$ stands for the exponent of
a trace of a matrix, and $\Gamma_p(.)$ denotes the multivariate
gamma function. It is also assumed that the innovation sequence
$\{\epsilon_t\}$ is uncorrelated and that $\{\epsilon_t\}$ is
uncorrelated with $\Sigma_0$, i.e. $\E(\epsilon_t\epsilon_s')=0$
(for any $t\neq s$) and $\E(\epsilon_t\textrm{vecp}(\Sigma_0)')=0$
(for all $t$). From the above inverted Wishart prior it turns out
that $\Sigma_0^{-1}$ follows the Wishart distribution with $n+p-1$
degrees of freedom and scale matrix $S_0^{-1}$, i.e.
$\Sigma_0^{-1}\sim W_p(n+p-1,S_0^{-1})$.

In order to completely specify this model, a value for the parameter
$k$ has to be specified. In Section \ref{model:estimation} it is
shown that in order to guarantee the expectation invariance property
(\ref{eq:mean1}) of the RW model, it is necessary to specify $k$ as
\begin{equation}\label{eq:k}
k=\frac{\delta(1-p)+p}{\delta(2-p)+p-1}.
\end{equation}

\subsection{Estimation}\label{model:estimation}

Suppose that at time $t$, the posterior distribution of $\Sigma_t$
is
\begin{equation}\label{post:t}
\Sigma_t|y^t \sim IW_p(n+2p, S_t),
\end{equation}
where $n=1/(1-\delta)$ and $S_t$ is known. For the singular multivariate beta density of $B_{t+1}$, we write
$B_{t+1}\sim B_p(m/2,1/2)$, where $m=\delta (1-\delta)^{-1}+p-1$. The ``singularity'' of the distribution derives from
$1<p-1$, for any $p>1$ and so the matrix $I_p-B_{t+1}$ is singular (for more details the reader is referred to
\cite{Uhlig94} and \cite{Diaz97}). The choice of $m$ is conveniently made so that two of the assumptions of the beta
density are satisfied, that is $m>p-1$ and $(1-\delta)n$ has to be an integer (see also the last paragraph of Section
\ref{model:estimation}).

Since $\Sigma_t^{-1}|y^t\sim W_p(n+p-1,S_t^{-1})$, from the
evolution (\ref{model2}) and from \cite{Uhlig94}, it follows that
$k^{-1}\Sigma_{t+1}^{-1}|y^t\sim W_p(n+p-1, S_t^{-1})$ or
$\Sigma_{t+1}^{-1}|y^t\sim W_p(n+p-1, kS_t^{-1})$ and so the prior
distribution of $\Sigma_{t+1}$ is
\begin{equation}\label{prior:t+1}
\Sigma_{t+1}|y^t \sim IW_p(\delta n+2p, k^{-1}S_t).
\end{equation}
From (\ref{post:t}) we have $\E(\Sigma_t^{-1}|y^t)=(n+p-1)S_t^{-1}$
and from (\ref{prior:t+1}) we have
$\E(\Sigma_{t+1}^{-1}|y^t)=(\delta n+p-1)kS_t^{-1}$, and so by
equalizing these two expectations we obtain
$$
k=\frac{n+p-1}{\delta n+p-1}=\frac{\delta(1-p)+p}{\delta(2-p)+p-1},
$$
as in (\ref{eq:k}). Using properties of the Wishart distribution, and adopting $k$ as proposed above, one can verify
that $\var(\textrm{vecp}(\Sigma_{t+1}^{-1})|y^t)\geq \var(\textrm{vecp}(\Sigma_t^{-1})|y^t)$, thus the RW type
evolution (\ref{model2}) is verified.

Proceeding now with the posterior distribution at time $t+1$, we
apply Bayes theorem by noting that the likelihood function from the
single observation $y_{t+1}$ is $p(y_{t+1}|\Sigma_{t+1})$, which
from (\ref{model1}) is the $p$-variate Gaussian distribution
$N_p(0,\Sigma_{t+1})$. Thus
\begin{eqnarray*}
p(\Sigma_{t+1}|y^{t+1}) &=& \frac{ p(y_{t+1}|\Sigma_{t+1},y^t)
p(\Sigma_{t+1}|y^t)}{ p(y_{t+1}|y^t) } \\ &\propto & \frac{
\textrm{etr}(-y_{t+1}'\Sigma_{t+1}^{-1}y_{t+1}/2)
|k^{-1}S_t|^{(\delta n+p-1)/2}
\textrm{etr}(-k^{-1}S_t\Sigma_{t+1}^{-1}/2) } {|\Sigma_{t+1}|^{1/2}
|\Sigma_{t+1}|^{(\delta n+2p)/2} } \\ &=& |\Sigma_{t+1} |^{-(\delta
n+1+2p)/2}
\textrm{etr}(-(y_{t+1}y_{t+1}'+k^{-1}S_t)\Sigma_{t+1}^{-1}/2),
\end{eqnarray*}
which is proportional to
\begin{equation}\label{post:t+1}
\Sigma_{t+1}|y^{t+1}\sim IW_p(n+2p,S_{t+1}),
\end{equation}
where $S_{t+1}=k^{-1}S_t+y_{t+1}y_{t+1}'$, since $\delta n+1=n$.

Equations (\ref{post:t}), (\ref{prior:t+1}) and (\ref{post:t+1}),
together with the prior (\ref{prior}) constitute a full algorithm,
for $t=1,\ldots,N-1$. We remark that, for $p=1$ and $k=1/\delta$,
the above results reduce to the usual algorithm for univariate
stochastic volatility estimation, as reported in \cite{West01} and
\cite{Triantafyllopoulos07}.

For $p\geq 1$, we see that, since $\delta<1$, we have
$\delta(1-p)+p>\delta(2-p)+p-1$ and so $0<k^{-1}<1$. Thus by
expanding $S_t$ as
$$
S_t=k^{-t}S_0+\sum_{j=1}^t k^{j-t}y_jy_j', \quad t=1,\ldots,N,
$$
we can approximate $S_t$ by
\begin{equation}\label{eq:St}
S_t\approx \sum_{j=1}^t k^{j-t}y_jy_j'
\end{equation}
and exclude the influence of the prior $S_0$, which anyway is
deflated as $t$ increases. We note that $S_t$ is just a weighted
average of the log-returns $\{y_jy_j'\}_{j=1,\ldots,t}$ with
weights $k^{-1}$. From this it follows that even if
$\{\Sigma_t^{-1}\}$ follows a random walk, the estimator $S_t$ is
still capable of exploiting mean reversion of the log-returns (as
it is a weighted average of the squares of log-returns) and thus
it is a suitable estimator for the volatility. The posterior mean
of $\Sigma_t$ and the prior mean at $\Sigma_{t+1}$ can be derived
easily from the inverted Wishart densities, i.e.
$$
\E(\Sigma_t|y^t)=\frac{S_t}{n-2}=\frac{(1-\delta)S_t}{2\delta-1}\quad
\textrm{and} \quad \E(\Sigma_{t+1}|y^t)=\frac{k^{-1}S_t}{\delta n-2}
= \frac{(1-\delta)S_t}{k(3\delta-2)},
$$
the posterior mean being defined for $\delta>1/2$ and the prior mean
being defined for $\delta>2/3$.

In related work, a number of authors such as \cite{Quintana1},
\citet[Chapter 16]{West01}, \cite{Aguilar00}, \cite{Liu},
\cite{Soyer06}, and \cite{Carvalho1} have suggested to use
$k=1/\delta$. Although it is easily verified that this is a correct
choice when $p=1$, setting $k=1/\delta$ when $p>1$ results in a
shrinkage-type evolution for $\Sigma_{t}^{-1}$. This can be seen by
first noting that, with $k=1/\delta$, we have
\begin{equation}\label{eq:ev:dif}
\E(\Sigma_{t+1}^{-1}|y^{t})-\E(\Sigma_{t}^{-1}|y^{t})=(p-1)(\delta^{-1}-1)S_{t}^{-1}
\end{equation}
and therefore the expectation is not preserved from time $t$ to
$t+1$, as we have
$\E(\Sigma_{t+1}^{-1}|y^{t})>\E(\Sigma_{t}^{-1}|y^{t})$.

In particular, when $p$ is large, even if $\delta\approx 1$, the above model postulates that the estimate of
$\Sigma_{t+1}^{-1}$ is larger than that of $\Sigma_{t}^{-1}$. In other words $\{\Sigma_t^{-1}\}$ follows an AR model
$\Sigma_t^{-1}=\alpha\Sigma_{t-1}^{-1}+\Gamma_t$, where $\alpha>1$; such a setting is clearly inappropriate. With the
RW type evolution of $\Sigma_t^{-1}$, claimed in all the above references, assuming that the limit of $S_t$ exists, it
follows from (\ref{eq:ev:dif}) that $0=(p-1)(\delta^{-1}-1)\lim_{t\rightarrow\infty}S_t$. This, for $p>1$, implies
that $\delta=1$ or $\lim_{t\rightarrow\infty}S_t^{-1}=0$, two meaningless results. Our suggestion is that $\delta$
should be replaced by $k^{-1}$, as in (\ref{eq:k}), a choice that now preserves the expectations.

Furthermore, for $p>1$, the updating equation of the degrees of
freedom of the Wishart distribution suggested in the above
references, namely
$$n_t+2p=\delta
n_{t-1}+1+2p=n_0\delta^t+(1-\delta^t)/(1-\delta)+2p,$$ does not seem
to be correct. The reason for this lies in the multivariate singular
beta distribution, $B_p(m_1/2,m_2/2)$ which is only defined for
$m_2$ being a positive integer \citep{Uhlig94}. Setting
$m_2=(1-\delta)n_t$, as in \cite{West01} and \cite{Soyer06}, results
in $m_2$ not being a positive integer. In our algorithm, we resolve
this issue by setting $n_t=n=1/(1-\delta)$ so that
$m_2=(1-\delta)n=1$. For more details on the multivariate singular
beta distribution the reader is referred to \cite{Uhlig94},
\cite{Diaz97}, and \cite{Srivastava03}; the density function of this
distribution is given in equation (\ref{Bt:pdf}) of the appendix.

\subsection{Performance measures}\label{model:performance}

\subsubsection{The likelihood function}

One method of model judgement and model comparison is via the
likelihood function. In this section, first we derive the
likelihood of our model in closed form. Adopting approximation
(\ref{eq:St}), the only parameters that need to be selected in
order to fully specify the model is the scalar $\delta$, since $k$
is specified in \eqref{eq:k}. Using the following result of
Theorem \ref{th1}, one possibility is to choose the value of
$\delta$ that maximizes the log-likelihood function (under the
restriction $2/3<\delta<1$).

\begin{thm}\label{th1}
In model (\ref{model1})-(\ref{model2}) the log-likelihood function
of $\Sigma_1,\ldots,\Sigma_N$, based on data $y_1,\ldots,y_N$ is
$$
c - \frac{1}{2}\sum_{t=1}^Ny_t'\Sigma_t^{-1}y_t +
\frac{2\delta-1}{2(1-\delta)}\sum_{t=1}^N\log |\Sigma_{t-1}|
  - \frac{p}{2} \sum_{t=1}^N\log |L_t|
-\frac{3\delta-2}{2(1-\delta)}\sum_{t=1}^N\log |\Sigma_t|,
$$
for
$$
c= -\frac{Np}{2}\log\pi - \frac{N}{2}\log 2\pi -
\frac{Np(2\delta-1)}{2(1-\delta)}\log k + N\log \frac{
\Gamma_p\{2^{-1}(1-\delta)^{-1}(\delta(1-p)+p)\} }{ \Gamma_p \{
2^{-1}(1-\delta)^{-1}(\delta(2-p)+p-1)\} },
$$
where $\delta>2/3$, $k$ is as in (\ref{eq:k}) and $L_t$ is the
diagonal matrix with diagonal elements the positive eigenvalues of
$I_p-k^{-1}\{\U(\Sigma_{t-1}^{-1})'\}^{-1}\Sigma_t^{-1}\{\U(\Sigma_{t-1}^{-1})\}^{-1}$,
with $\Sigma_t^{-1}=\U(\Sigma_t^{-1})'\U(\Sigma_t^{-1})$.
\end{thm}

The proof of this result can be found in the appendix. A common modelling strategy in Bayesian inference is to plug
the posterior mean of $\Sigma_t$ in to the likelihood function and then to compare models by comparing their
likelihood functions (e.g. see \cite{Leonard99}). This approach has common roots to estimation methods using the
profile likelihood \citep{Lutkepohl,Leonard99}, and clearly it has the advantage of combining Bayes estimation with
likelihood-based inference. In addition to that, this approach can be very useful for choosing nuisance parameters,
such as the discount factor $\delta$. The maximization of the log-likelihood function with respect to $\delta$ may be
slow because this is a non-linear function in $\delta$. A possibility would be to evaluate the log-likelihood function
only on a few admissible values for $\delta$ $(2/3<\delta<1)$. Values of $\delta$ lower than $0.7$ can result in very
volatile, not smooth, and thus unstable posterior estimates of $\Sigma_t$; values of $\delta$ larger than $0.95$ can
result in very smooth estimates of $\Sigma_t$, not able to capture the clusters and the spikes of the volatility. In
this paper (see the illustration of Section \ref{example:fx}), we recommend exploring values of $\delta$ in the range
$0.7,0.75,0.8,0.85,0.9,0.95$. \cite{West01} and \cite{TRGPN07} have some discussion on the performance of the
posterior estimates at the boundary values of discount factors $\delta>0.95$.

\subsubsection{One-step forecast error}

Other than the log-likelihood function, the mean of square
standardized one-step forecast error vector (MSSE) provides
another performance measure. From (\ref{model1}) the one-step
forecast distribution of $y_{t+1}|y^t$ is a $p$-variate Student
$t$ density with $\delta/(1-\delta)$ degrees of freedom, mean
vector 0 and scale matrix $k^{-1}S_t$, written $y_{t+1}|y^t\sim
t_p(\delta/(1-\delta),0,k^{-1}S_t)$ \citep{Gupta}. It then follows
that, for $\delta>2/3$,
$$
\var(y_{t+1}|y^t)=\frac{k^{-1}S_t}{\delta
/(1-\delta)-2}=\frac{(1-\delta)S_t}{(3\delta-2)k},
$$
which also can be derived from Section \ref{model:estimation}, using
conditional expectations, i.e.
$$
\var(y_{t+1}|y^t) = \E ( \var (y_{t+1}|\Sigma_{t+1},y^t)|y^t) = \E
(\Sigma_{t+1}|y^t)=\frac{(1-\delta)S_t}{(3\delta-2)k},
$$
since from model (\ref{model1}), it is
$\var(\E(y_{t+1}|\Sigma_{t+1},y^t)|y^t)=0$. Having obtained an
expression for the variance, we can now write the standardized
one-step forecast error vector $u_{t+1}$ as
\begin{equation}\label{eq:se:pdf}
u_{t+1}=\sqrt{k}S_t^{-1/2}y_{t+1}\quad \textrm{with} \quad
u_{t+1}|y^t\sim t_p\left(\frac{\delta}{1-\delta},0,I_p\right)
\end{equation}
so that the vector
$$
u_{t+1}^*=\left\{\frac{(1-\delta)S_t}{(3\delta-2)k}\right\}^{-1/2}y_{t+1}
$$
has $\E(u_{t+1}^*|y^t)=0$ and $\E(u_{t+1}^*(u_{t+1}^*)'|y^t)=I_p$.
Then the MSSE vector is given by
\begin{gather*}
\textrm{MSSE} = \frac{1}{N} \sum_{t=1}^N \left\{(u_{1t}^*)^2,
\ldots, (u_{pt}^*)^2 \right\}',
\end{gather*}
where $u_t^*=(u_{1t}^*,\ldots,u_{pt}^*)'$. Models that fit well the
data are expected to yield $\textrm{MSSE}\approx(1,\ldots,1)'$.

\subsubsection{Bayes factors}

A third approach for model diagnostics is based on sequential Bayes
factors \citep{West01, Salvador1, triantafyllopoulos06}. Suppose we
have two competing models, $\mathcal{M}_1$ and $\mathcal{M}_2$,
parameterized in terms of $\delta_1$ and $\delta_2$, respectively.
First, a Bayes factor is obtained as the logarithm of the ratio
between the density of $u_t\equiv u_t(\delta_1)$ (under
$\mathcal{M}_1$) and the density of $u_t\equiv u_t(\delta_2)$ (under
$\mathcal{M}_2$). Specifically, at each time $t$ we have
$$
H_t = \log \frac{ p(u_t(\delta_1)|y^{t-1},\mathcal{M}_1) } {
p(u_t(\delta_2)|y^{t-1},\mathcal{M}_2) }, \quad t=1,\ldots,N,
$$
and, from the Student $t$ density (\ref{eq:se:pdf}), this becomes
$$
H_t = \frac{ \Gamma((n_1+p)/2) \Gamma(n_2/2) }{ \Gamma((n_2+p)/2)
\Gamma(n_1/2) } \left\{ \frac{
(1+u_t(\delta_2)'u_t(\delta_2))^{n_2+p} }{
(1+u_t(\delta_1)'u_t(\delta_1))^{n_1+p}} \right\}^{1/2},
$$
where $\Gamma(.)$ denotes the gamma function and
$n_i=\delta_i/(1-\delta_i)$, for $i=1,2$.

A value of $H_t>0$ then suggests that model $\mathcal{M}_1$ has to
be preferred over $\mathcal{M}_2$, in the sense that $\mathcal{M}_1$
is associated with a superior forecast distribution. Alternative,
negative values for $H_t$ suggest that $\mathcal{M}_2$ is the
preferred model. In situation where $H_t=0$, both models are deemed
equivalent. One point of interest is what decision can we make when
$H_t$ fluctuates around zero. In such a case one may select a
threshold value in order to decide which model to choose, as in
\cite{West01}.

\section{An illustration using foreign exchange rates}\label{example:fx}

\begin{figure}[t]
\begin{center}
 \epsfig{file=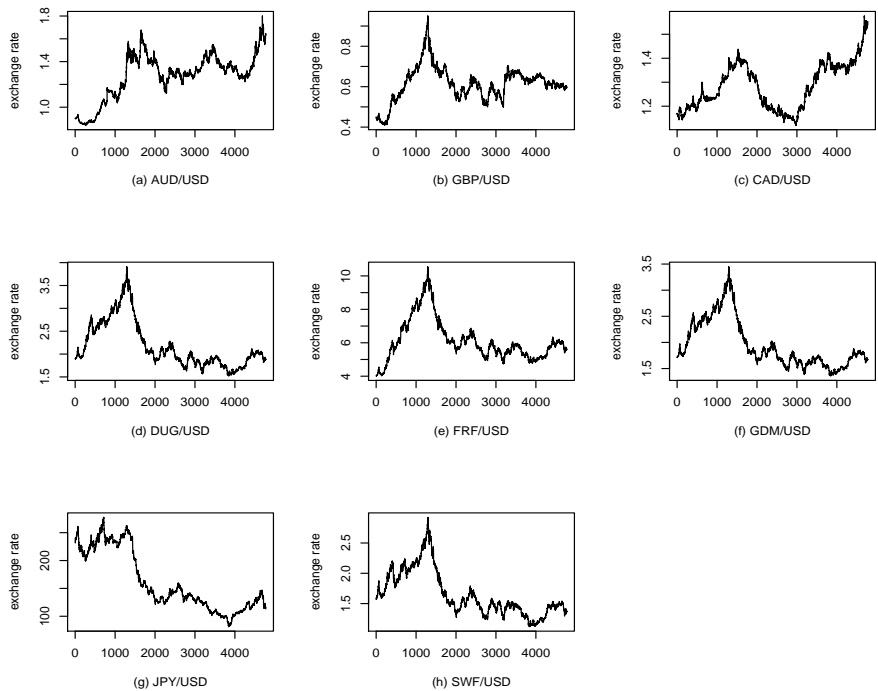, height=10cm, width=12cm}
 \caption{Daily observations on eight foreign exchange rates.}\label{fig1}
 \end{center}
\end{figure}

\begin{figure}
\begin{center}
 \epsfig{file=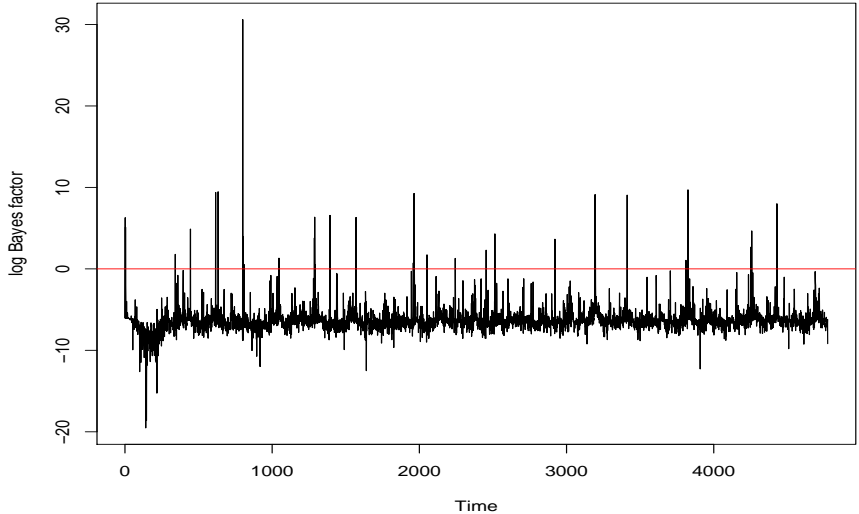, height=8cm, width=12cm}
 \caption{Sequential Bayes factor $H_t$ of the standardized one-step forecast errors of model
 $\mathcal{M}_1$ $(\delta=0.7)$ vs model $\mathcal{M}_2$ $(\delta=0.95)$.}\label{fig2}
 \end{center}
\end{figure}

In this section we present an analysis of eight exchange rates
\emph{vis-$\grave{a}$-vis} the US dollar. The exchange rates are the
Australian dollar (AUS), British pounds (GBP), Canadian dollar
(CAD), German Deutschmark (GDM), Dutch guilder (DUG), French frank
(FRF), Japanese yen (JPY) and Swiss franc (SWF), all expressed as
number of units of the foreign currency per US dollar. The sample
period runs from $2$ January $1980$ until $31$ December $1997$, and
corresponds to $4774$ observations, sampled at daily frequencies.
This data set was originally obtained from the New York Federal
Reserve, and then discussed in \cite{Franses}. Figure \ref{fig1}
illustrates the daily observations on the level of all eight
exchange rates.

We have applied the stochastic volatility model of Section
\ref{model:estimation} to the logarithmic returns, which have been
collected in a vector $y_t=(y_{1t},\ldots,y_{8t})'$. Following the
empirical studies of exchange rates, as in \cite{Quintana1},
\cite{Putnam}, and \cite{Quintana2}, we adopt the random walk for
the evolution of the volatility and thus we specify $k$ as in
(\ref{eq:k}). In order to choose a suitable value for the parameter
$\delta$, we have used the performance measures described in Section
\ref{model:performance}. Following suggestions in that section, we
have only considered a few selected values of $\delta$ in the range
$0.7\leq\delta\leq 0.95$. The results from this analysis are
summarized in Table \ref{table1}, which provides the mean of the
MSSE (MMSSE), the log-likelihood function (evaluated at the
posterior mean of the volatility), and the mean of the Bayes factors
of the standardized one-step forecast errors. For the computation of
the Bayes factors here, each $\mathcal{M}_1$ is based on the current
value of $\delta$, and is compared against a baseline model
$\mathcal{M}_2$ that uses $\delta=0.95$.

From Table \ref{table1}, it can be observed that for small values of
$\delta$, the MMSSE also attains small values, indicating poor
performance, when compared to an ideal MMSSE value of one. This
result seems to suggest that the forecast covariance matrix of $y_t$
has been over-estimated. As $\delta$ gets close to one, the MMSSE
also gets close to one, which underlines an improvement in the
estimation of the forecast covariance matrix of $y_t$. The
log-likelihood function attains its largest value at $\delta=0.95$.
For each $\delta<0.95$, the Bayes factor mean $H$ is negative and
this indicates a preference in favour of model $\mathcal{M}_2$ (for
$\delta=0.95$). In particular we note that the model performance
deteriorates as $\delta$ decreases, a fact that is captured by all
three diagnostic measures considered here. As a result of this, we
conclude that $\delta=0.95$ produces the best model.

Figure \ref{fig2} shows the log-Bayes factor sequence $\{H_t\}$,
from which the superiority of model $\mathcal{M}_2$ is clear. We
observe that, out of $N=4774$ data points, $\{H_t\}$ is positive at
only $37$ points (i.e. only $0.77\%$ of the time). Using sequential
Bayes factors, the modeler has the extra advantage of choosing the
discount factor at each time $t$ according to the sign of $H_t$.
This is particularly advantageous in an on-line setting, and when
decisions have to be made in real time.

\begin{table}
\caption{Mean (over the eight exchange rates) of the mean square
one-step forecast standardized errors (MMSSE), log-likelihood
function (LogL) evaluated at the posterior mean of the volatility,
and mean of the log Bayes factor $H_t$
$(t=1,\ldots,4774)$.}\label{table1}
\begin{center}
\begin{tabular}{c|ccc}
\hline $\delta$ & MMSSE & LogL & $H$  \\ \hline 0.70 & 0.072 &
-12857.59 & -6.269   \\ 0.75 & 0.194 & -12395.30 & -5.681   \\
0.80 &
0.337 & -11721.93 & -4.950 \\ 0.85 & 0.506 & -10644.32 & -3.982  \\
0.90 & 0.701 & -8627.03 & -2.564
\\ 0.95 & 0.912 & -3458.23 & 0  \\
\hline
\end{tabular}
\end{center}
\end{table}

\begin{figure}[t]
\begin{center}
 \epsfig{file=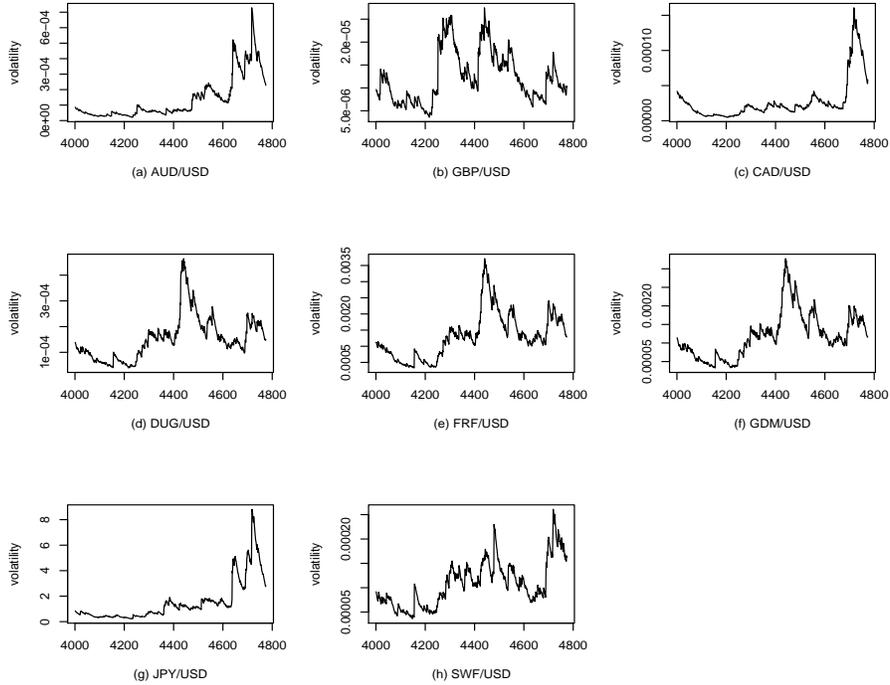, height=10cm, width=12cm}
 \caption{Estimate of the posterior volatility for the FX data,
 using the model with $\delta=0.95$.}\label{fig3}
 \end{center}
\end{figure}

\begin{figure}[t]
\begin{center}
 \epsfig{file=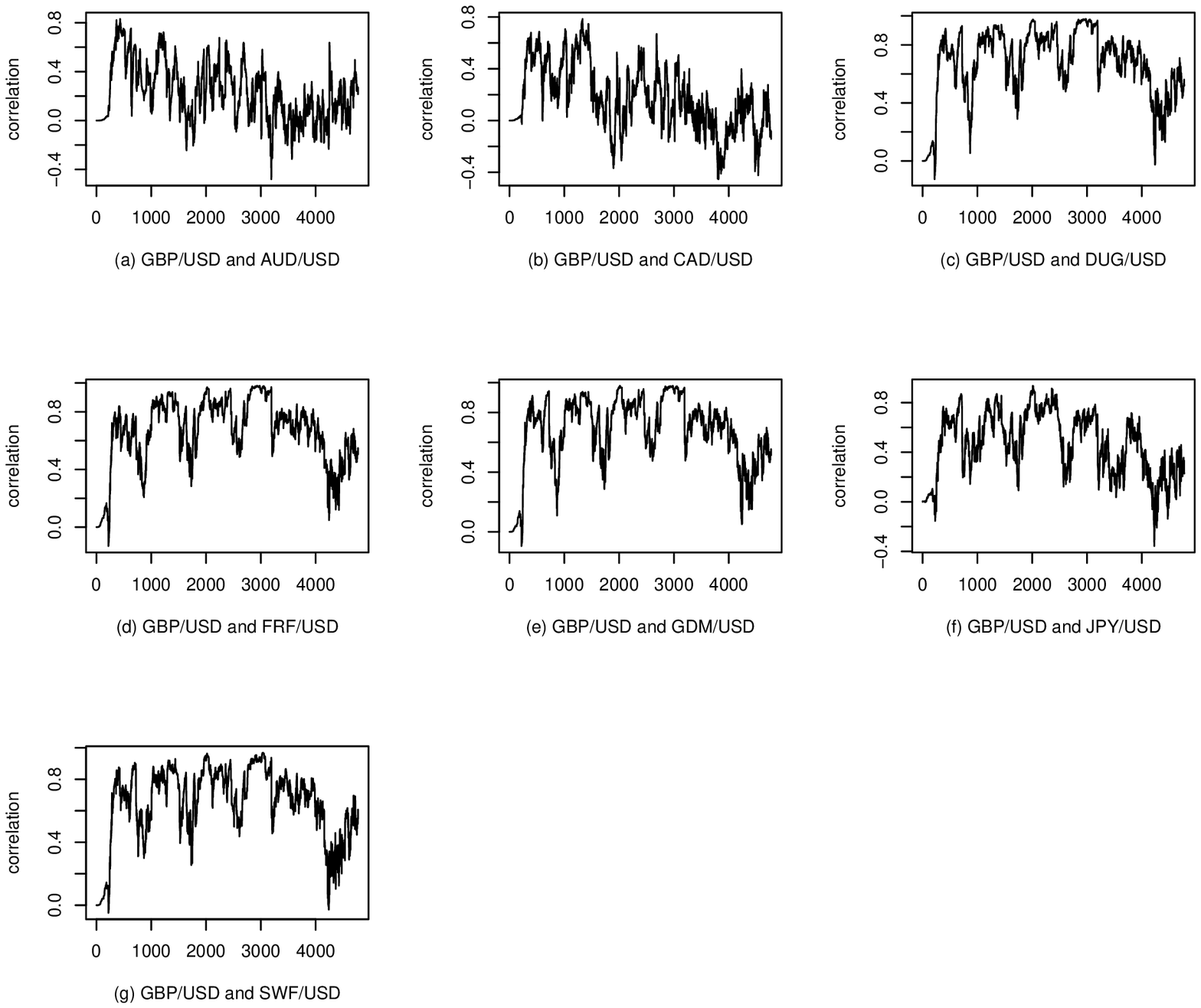, height=10cm, width=12cm}
 \caption{Estimate of the posterior correlation coefficient for
 the exchange rates data using the model with $\delta=0.95$.}\label{fig4}
 \end{center}
\end{figure}

Figure \ref{fig3} shows the posterior volatilities, i.e. the
estimates of $\sigma_{ii,t}$ $(i=1,\ldots,8)$, for a subset of the
data points ($t=4001,...,4774$). Most of the volatilities are small,
except for the JPY/USD; even for small volatilities, this figure
indicates clearly the highly volatile periods for each exchange
rate. Figure \ref{fig4} shows the posterior correlations of GBP/USD
versus all the other rates. This figure confirms that the
correlations are time-varying. By inspecting Figure \ref{fig4} we
observe that GBP/USD is most correlated with DUG/USD, FRF/USD,
GDM/USD, and SWF/USD.

Finally we note that, for this relatively large data set, based on
$4774$ time points in $8$ dimensions, the estimation algorithm
(implemented in the {\tt R} language on a Windows platform) took
less than a minute ($55$ seconds) to complete, on a PC with Intel(R)
Celeron(R)M Processor 1.60GHz and 504MB RAM, including the
evaluation of the log-likelihood function and the Bayes factors.

\section{Conclusions}\label{discussion}

In this paper we have described a Bayesian modeling approach for
multivariate stochastic volatility. The proposed estimation
methodology is delivered in closed form, is easily implementable and
efficient, as the model relies on only one parameter.

The models proposed in this paper are closely related to the above
mentioned articles as well as to the models of \cite{Uhlig97} and
\cite{Philipov2}. Notably, we have shown that similar volatility
estimators proposed in the literature are based on a shrinkage-type
volatility evolution, which is not a realistic choice. Instead, the
estimator described here guarantees a random walk type evolution.

The procedure proposed in this paper attempts to combine the simplicity of non-iterative algorithms with the
sophistication of stochastic volatility models. In our view, algorithms such as the one suggested here are
particularly attractive because they can model high dimensional data with low computational cost, which is crucial for
certain real-time applications in modern computational finance, such as algorithmic trading. Future research efforts
will be directed towards other financial applications with special focus on optimal portfolio allocation.

\renewcommand{\theequation}{A-\arabic{equation}} 
\setcounter{equation}{0}  
\section*{Appendix}

\begin{proof}[Proof of Theorem \ref{th1}]
First we derive the density of $\Sigma_t|\Sigma_{t-1}$, for
$t=1,\ldots,N$. From (\ref{model2}), it is $B_t\sim B_p(m/2,1/2)$,
for $m=\delta(1-\delta)^{-1}+p-1$, with density
\begin{equation}\label{Bt:pdf}
p(B_t)=\pi^{-p/2} \frac{ \Gamma_p((m+1)/2)}{\Gamma_p(m/2)}
|K_t|^{-p/2}|B_t|^{(m-p-1)/2},
\end{equation}
where $I_p-B_t=H_1K_tH_1'$, $K_t$ is the diagonal matrix with
diagonal elements the positive eigenvalues of $I_p-B_t$, and $H_1$
is a matrix with orthogonal columns, i.e. $H_1H_1'=I_p$. For more
details on this distribution see \cite{Uhlig94}.

Now from evolution (\ref{model2}) we have the transformation from
$B_t$ to $\Sigma_t=k^{-1}(\U(\Sigma_{t-1}^{-1}))^{-1} B_t^{-1}$
$\times (\U(\Sigma_{t-1}^{-1})')^{-1}$. From \cite{Diaz97} the
Jacobian of this transformation is
$$
(\,d B_t)=|K_t|^{p/2} |L_t|^{-p/2} k^{-p/2} |\Sigma_{t-1}|^{1/2}
(\,d\Sigma_t),
$$
where $L_t$ is the diagonal matrix with diagonal elements the
positive eigenvalues of $I_p-k^{-1}(\U(\Sigma_{t-1}^{-1})')^{-1}
\Sigma_t^{-1} (\U(\Sigma_{t-1}^{-1}))^{-1}$. From the above
transformation it is
$$
|\U(\Sigma_{t-1}^{-1})|=|\U(\Sigma_{t-1}^{-1})'\U(\Sigma_{t-1}^{-1})|^{1/2}=
|\Sigma_{t-1}|^{1/2} \quad \textrm{and} \quad
|B_t|=k^{-p}|\Sigma_{t-1}| |\Sigma_t|^{-1}
$$
and thus from (\ref{Bt:pdf})
\begin{eqnarray}
p(\Sigma_t|\Sigma_{t-1}) &=& \pi^{-p/2} \frac{
\Gamma_p((m+1)/2)}{\Gamma_p(m/2)} k^{-p/2} |\Sigma_{t-1}|^{1/2}
\nonumber \\ && \times |K_t|^{-p/2} |B_t|^{(m-p-1)/2} |K_t|^{p/2}
|L_t|^{-p/2} \nonumber \\ &=& \pi^{-p/2} k^{-p(m-p)/2} \frac{
\Gamma_p((m+1)/2)}{\Gamma_p(m/2)} |L_t|^{-p/2} \nonumber \\ &&
\times |\Sigma_{t-1}|^{(m-p)/2} |\Sigma_t|^{-(m-p-1)/2}.
\label{app:sigma}
\end{eqnarray}
For the likelihood function $L(\Sigma;y)$, where
$\Sigma=(\Sigma_1,\ldots,\Sigma_N)$ and $y=(y_1,\ldots,y_N)$, write
$$
L(\Sigma;y)=\prod_{t=1}^Np(y_t|\Sigma_t)p(\Sigma_t|\Sigma_{t-1}).
$$
From equation (\ref{model1}) we have $y_t|\Sigma_t\sim
N_p(0,\Sigma_t)$, while the density of $\Sigma_t|\Sigma_{t-1}$ is
given by (\ref{app:sigma}). The required formula of the
log-likelihood function is obtained by taking the logarithm of
$L(\Sigma;y)$.
\end{proof}


\bibliographystyle{plainnat}
\bibliography{bibliography}

\end{document}